\documentstyle[psbox]{WORKSHOP}
\begin{document}

\newcommand{\KU}{\mbox{\rm 4U 1630-472}}

\title{
Disk emission and absorption lines in LMXB. \\ 
Note on the physical conditions  of an absorbing material
} 

\author{
Agata R\'o\.za\'nska$^1$
\\[12pt]  
%
$^1$  N. Copernicus Astronomical Center, Bartycka 18, 00-716 Warsaw, Poland\\
%
{\it E-mail(AR):agata@camk.edu.pl } 
}

\abst{
X-ray binaries often show continuum spectrum that is 
interpreted correctly as emission from an accretion disk.
Additionally,  two absorption lines from He and H-like iron have been clearly 
detected  in  the high-resolution data from several such sources,
for instance \KU\ . We show that the continuum X-ray spectrum of \KU\ 
 with iron absorption lines can be satisfactorily modeled by the 
spectrum from an accretion disk atmosphere.
We performed full radiative transfer calculations using our code ATM21 to model 
the emission from an accretion disk surface that is seen at 
different viewing angles. Computed models 
are then fitted to  the high-resolution X-ray spectra of \KU\ obtained by  {\it Suzaku}
satellite. Absorption lines of highly ionized iron originating in a hot 
accretion-disk atmospheres are important part of the observed line profile,
and can be an alternative or complementary
explanation to the wind model usually  favored for this type of sources.

Next, assuming that absorption lines originate from the wind 
illuminated by X-ray central source in LMXBs,  we can put constrains on the wind 
location only if we know the volume density number of the absorbing material. 
There are a few derivations of the distance to the wind in X-ray binaries. 
We show here, that the  density number and wind location agree with density 
 of an upper disk atmosphere at optical depth, $\tau=2/3$, at the same distance 
from the black hole.
This comparison is done assuming optically thick, geometrically thin standard 
accretion disk model. Nevertheless, it shows that the wind physical conditions 
are the same as in thermalized disk gas, and we only have to figure out 
how the wind is blowing?
}

\kword{X-rays: binaries, Stars: individual, Accretion disks, 
          Radiative transfer, Line: profiles}

\maketitle
\thispagestyle{empty}

\section{LMXBs with absorption lines}

The Low Mass X-ray Binaries (LMXBs) are binary accreting systems, where 
the mass flows onto the neutron star (NS) or the black hole (BH) from the companion. 
The Secondary is usually main sequence star of the late type
K or M  that is of the low weight.  

When the magnetic field of such a system is negligible, flowing matter with a certain 
angular momentum can form accretion disk around compact object.
According to any global disk models considered for the mass of 
central object that is close to 10 $M_{\odot}$ or less, the effective
temperature of the inner radii reaches $10^7$  K  (R\'o\.za\'nska et al. 2011).
The temperature at a given radius always increases with the 
decreasing mass of the central object and with the increasing accretion rate.
For such high temperatures of an accretion disk atmosphere, thermal lines 
from H-like and He-like iron ions are formed, and  
those lines should be visible between 6.7 to 9.2 keV, 
where the latter value is the  energy of the last iron bound-free transition 
(from H-like ion to complete ionization). Narrow lines created in the 
inner disk region are relativistically smeared, but if they are created
far enough, can survive relativistic motion. 

\begin{figure*}[t]
\centering
\psbox[xsize=8cm,rotate=r]{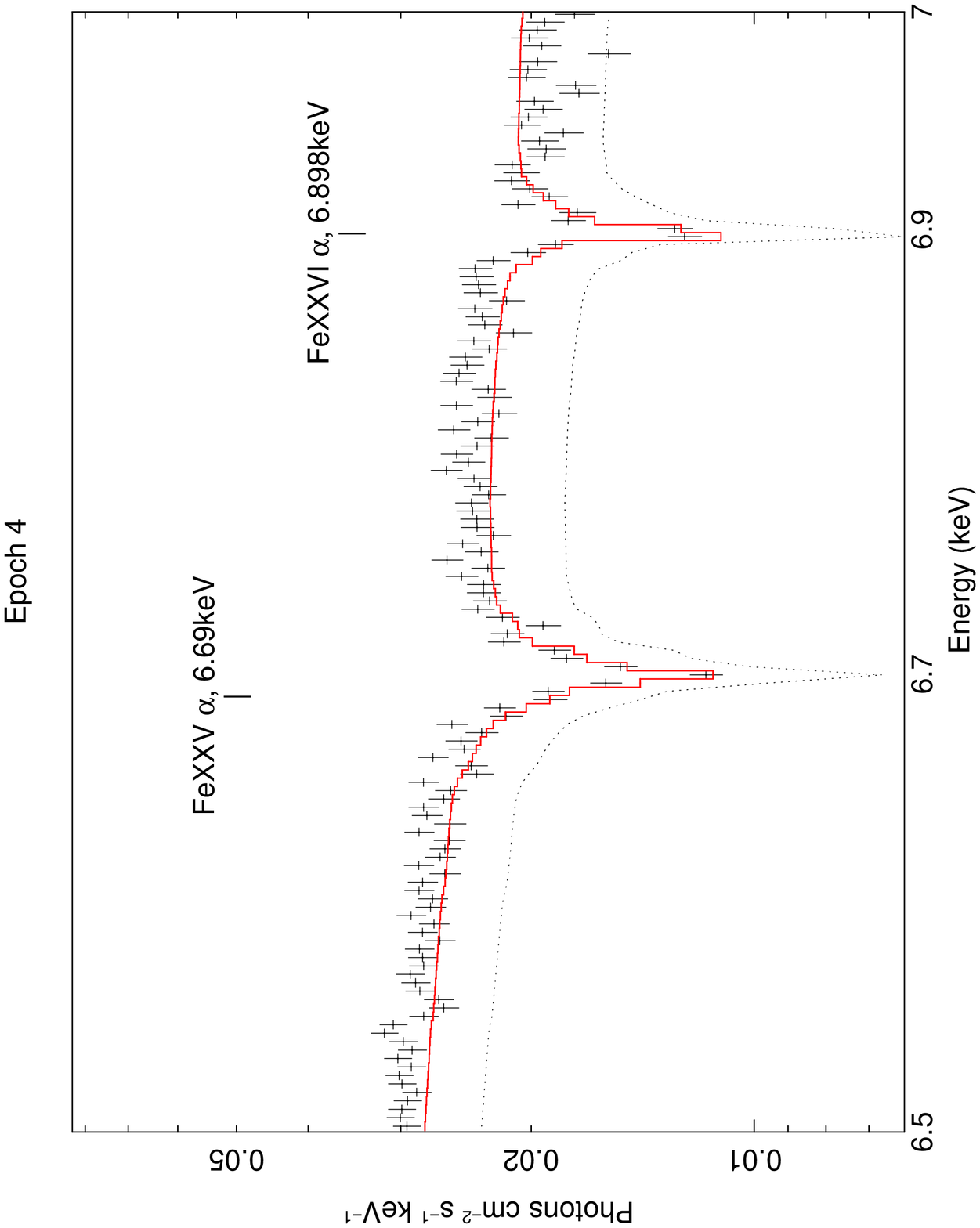}
\psbox[xsize=8cm,rotate=r]{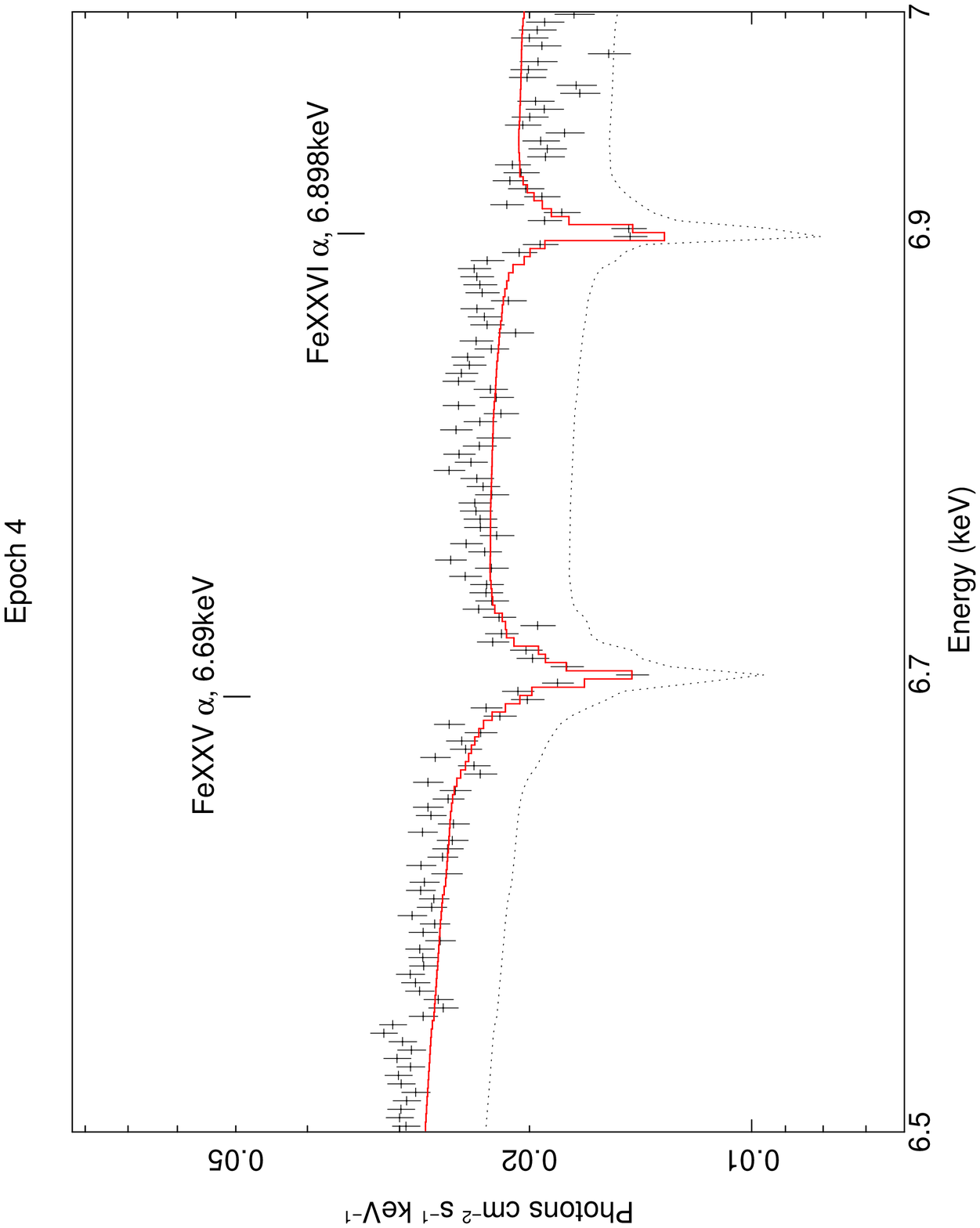}
\caption{Focus on the resonant He and H-like iron lines for the fitting done 
in the 6-9 keV energy range, fitted to the {\it Suzaku} data of \KU\ .
Black crosses show the data, black dotted lines represent the 
{\sc atm} atmospheric model, while red solid lines 
show the  total model. Clearly, complicated 
line profiles computed in our model, match the observations very well.
Left panel shows slightly better fit done for viewing angle  $i=11\pm 5^{\circ}$
with $\chi ^2/d.o.f. = 1.52$
while right panel shows fit for $i=70\pm 6^{\circ}$ with $\chi ^2/d.o.f. = 1.57$
}
\label{fig:obs}
\end{figure*}

Emission from the accretion disk around a compact object is the 
commonly accepted model for the soft X-ray bump observed in 
LMXBs. 
Nevertheless, it is well known that observed X-ray spectra from 
compact binaries do not always exhibit a disk component. 
Owing to  instrument's technical limits and to 
spectral state transition, multi-temperature disk component 
may not be detected in full X-ray energy range.
To explain data by the accretion disk emission, we have to be sure 
that observations were taken when the source was in the so-called 
``soft state'', which is dominated by disk-like component.

Several recently observed X-ray binaries have exhibited absorption 
lines from highly ionized iron (Borin et al. 2004, Kubota et al. 2007, 
Miller et al. 2008, D\'iaz Trigo et al. 2012).
Many of those sources show dips in their light curves, which are believed
to be caused by obscuration of the central X-ray source by a dense material 
located at the outer edge of an accretion disk. 
Such obscuring material was accumulated during the accretion phase from the 
companion star onto the disk (White \& Swank 1982). The presence of dips and
the lack of total X-ray eclipses by the companion star indicate that the 
system is viewed relatively close to edge-on, at an inclination angle in 
the range $\sim 60-80^{\circ} $ (Frank et al. 1987).

The He-like and H-like Fe  absorption features indicate that highly ionized
plasma is present in these systems. Study of these lines is extremely
important for characterizing the geometry and physical properties of plasma. 
Recently, it has been shown that the presence of absorption lines is not 
necessarily related to the viewing 
angle since non-dipping sources also show those features (D\'iaz Trigo et al. 2012). 
Moreover, the Fe{\sc xxv} absorption line was also observed during  
non-dipping intervals in XB1916-053 (Borin et al. 2004).
On the other hand, it was shown by Ponti et al. (2012) that the strength of such lines
depends on the spectral state of the X-ray binary.  
In this paper, we show that such absorption lines can be well fitted by 
a single disk model, where the line profile is computed taking into account all proper 
line broadenings (Sec.~\ref{sec:obs}). 

There is a growing number of indications that the density of winds in 
LMXBs is of the order of $10^{17}$ cm$^{-3}$ (D\'iaz Trigo et al. 2013, 
Miller et al. 2013). Additionally, this wind is located within $r=10^{10}$ cm.
Bellow, we calculate vertical disk structure (R\'o\.za\'nska et al. 1999) in the way 
commonly used for studying accretion disk stability curves (Smak 1983, Hameury \& Lasota 2005).
Assuming that disk is fully thermalized we show how the density number at $\tau=2/3$ of 
the disk gas depends on the distance from the central object. 
We show in Sec.~\ref{sec:mod}, that the observed wind physical conditions 
(Miller et al. 2013) coincide with 
density of an accretion disk atmosphere assuming standard geometrically 
thin, optically thick disk (Shakura \& Sunyaev 1973).  
Conclusions are given in Sec.~\ref{sec:con}

\section{The case of \KU\ }
\label{sec:obs}

Recently, we presented fitting of complex continuum and line numerical 
models to X-ray spectra of \KU\ (R\'o\.za\'nska et al. 2014).
In our models, the spectrum of disk emission 
was obtained from careful radiative transfer computations including Compton 
scattering on free electrons. The Fe line profiles are computed 
as the convolution of natural, thermal, and pressure broadening mechanisms. 
The advantage of our models 
is that the continuum is fitted with lines simultaneously, which has never 
been done before in the analysis of X-ray absorption lines seen in LMXBs. 
The usual procedure is to fit both the disk emission as a standard model 
in XSPEC fitting package, and Gaussian lines, where the energy of line 
centroid is a free parameter of fitting. In such a case, lines are usually 
blue-shifted indicating that absorbing matter outflows. 

The accretion-disk atmosphere spectra fit the  {\it Suzaku} data 
for \KU\ very well as presented at Fig.~\ref{fig:obs}. 
The best fit we have obtained for the inclination angle $i=11^{\circ}$
- left panel. 
For higher angles, i.e. $i=70^{\circ}$, the fit is just slightly worse - right panel. 
This angle is within the range of inclination suggested in the literature 
when taking  dipping behavior into account and assuming absorption
 in the wind. 
The small difference in fit quality between different inclinations does not 
allow us to claim constraints on inclination angle.

We modeled continuum and line spectra using a  single model. 
Absorption lines of highly ionized iron can originate 
in  the upper parts of the disk static atmosphere, which is intrinsically hot
because of the high disk temperature. Iron line profiles computed with 
natural, thermal, and pressure broadening match observations very well
(R\'o\.za\'nska et al. 2014).

In this work we do not aim to tightly constrain parameters of the object 
but rather to show that emission from the accretion disk atmosphere is an 
important mechanism that gives a vital explanation or at least part 
of an answer to the question of the origin of iron absorption in 
X-ray binaries.

The major conclusion of our analysis is that the shape of disk spectrum 
is interpreted well as \KU\ emission and that absorption lines do not need
to set any velocity shift to explain data. Therefore, the wind explanation
for absorbing matter is questionable and not unique. We showed that X-ray
data of current quality can be interpreted in several ways, and we cannot
easily solve this ambiguity.

\section{Connection of the wind with an accretion disk atmosphere}
\label{sec:mod}

The wind modelling of X-ray absorption has one major difficulty:
we need independent volume density measurement to determine the 
wind location according to the standard formula:
\begin{equation}
\xi = {L_{ion} \over n_{0}R^2} ,
\label{eq:ion}
\end{equation} 
where $L_{ion}$ is the wind ionizing luminosity, $n_{0}$ is the hydrogen 
number density at the wind illuminated surface, and $R$ is the distance of 
the absorber from an UV/X-ray source, i.e. inner disk, compact object or 
eventual X-ray corona.

Photoionzation models are degenerated, when we assume hard X-ray power-law 
as the spectral energy distribution (SED) of ionizing source. In such the case, 
transmitted spectrum from rare more distant cloud looks the same 
as from dense cloud located close to the center. Therefore, it is not possible to 
estimate volume density from photoionization calculations for hard 
power-low X-ray continuum.

Up to now, there are four independent methods of density diagnostic in the 
wind, but all of them work only in the particular range of parameters: i) variability
method (Krolik \& Kriss 1995), ii) measurement of the ratio of photoexcitation 
lines of Fe{\sc xxii} (Mauche et al. 2003)
iii) measurement of ionic column densities of the 
excited metastable states of low ionization ions C{\sc ii}, Fe{\sc ii} 
(Korista et al. 2008), and 
iv) photoionization modelling valid only if SED is dominated by soft UV/Soft-X-ray 
component
(R\'o\.za\'nska et al. 2008).

In the case of winds in LMXBs, where ionization is very high, the second method
is the most suitable, as it was shown recently by Miller et al. (2013) in case 
of MAXI J1305-704 {\it Chandra} data. 
The authors estimated gas density number to be higher than $10^{17}$ cm$^{-3}$. 
Together with the luminosity of the source to be 
$L=10^{37}$ erg/s and ionization parameter $log(\xi)=2.05$, it gives the wind location within 
$R=3.9 \times 10^8$ - $3 \times 10^9$ cm, depending on the fitted model
(see Tab.~5 in Miller et al. 2013 paper). 

In other sources when exited levels are not detected, 
 with Eq.~\ref{eq:ion} we can always put upper limit for 
the density assuming that the wind size is comparable to the distance from 
an ionizing source. It was done for several objects and summarized at 
Fig.~1 of paper by D\'iaz Trigo \& Boirin (2013). 

\begin{figure}
\centering
\psbox[xsize=8.5cm]{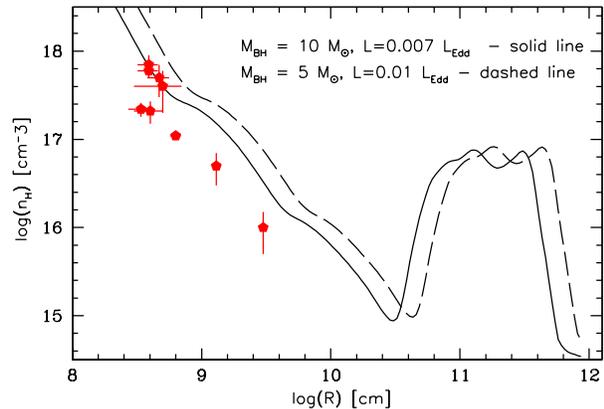}
\caption{Number density value at $\tau=2/3$ in accretion disk around
black hole of 10 and 5 Solar masses 
versus the distance from the black hole. The assumed accretion 
rates are 0.007 and 0.01 of Eddington accretion rate respectively. 
The density was calculated 
using full set of differential equations with parameters given in the 
text. Points mark the physical parameters of the wind found 
by Miller et al. 2013 in case of MAXI J1305-704 {\it Chandra} data.
We adopted points from models: 1,2,3,4,5 and 8 fitted to the data
and listed in Tab.~5 of this paper.}
\label{fig:mod}
\end{figure}

In this paper, we made calculations of accretion disk vertical structure 
solving standard differential equations for 1D fully thermalized 
gas (Pojma\'nski 1985, R\'o\.za\'nska et al 1999). Such calculations are extensively used to 
compute time evolution of accretion disk instabilities which 
explain outbursts presented in optical/X-ray data of accreting compact objects
(Smak 1983, Hameury \& Lasota 2005). 

For our purpose, we calculated vertical structure on different distances from 
the central black hole of the 5 and 10 Solar masses with accretion rates 0.01 and  
0.007  of Eddington unit, respectively. Such accretion rates are used 
in order to reproduce luminosity 
used by Miller et al. 2013 for distance determination. 
Additionally, we assume standard accretion efficiency 
for non-rotating black hole 1/16, and viscosity parameter $\alpha=0.1$.
The set of parameters is very basic, and we are aware that different disk 
model can change results, but up to know standard disk very well explains 
observations.  

At each distance from the black hole we determined the density number of fully 
thermalized, i.e. at $\tau=2/3$, gas. In Fig.~\ref{fig:mod} we present
radial dependence of such density for two cases of black hole masses and 
accretion rates pointed in the right upper corner. The radiative 
transfer in those calculations is solved by diffusion approximation,
where radiative flux is proportional to the temperature gradient and 
inversely proportional to the opacity of matter. All visible bumps are 
due to opacities which are taken to be Rosseland means, but self consistently 
computed for Solar abundances (R\'o\.za\'nska et al. 1999).  

Additionally, several points measured by Miller et al. (2013) are putted on the 
figure. We have taken models numbered: 1,2,3,4,5, and 8 from their paper listed
in Tab.~5.  Those points indicate wind physical conditions 
in MAXI J1305-704 determined by
fitting photoionized XSTAR model to the data. 
The fitting is done on the level where, ionic column 
densities of photoexited Fe{\sc xii} lines computed in the model are compared with 
those observed by {\it Chandra} X-ray telescope. 

One can see strong connection of the wind physical parameters
with eventual accretion disk atmosphere. This fact may suggest that the
wind consist the same material as in upper thermalized disk atmosphere.

\section{Discussion} 
\label{sec:con}

The high density of absorbing matter in LMXBs was first suggested by 
Frank et al. (1987), who proposed the two-phase 
medium to explain X-ray observations of such sources. 
The Authors have concluded that the absorbing material has 
density number of the order of $10^{16}$ cm$^{-3}$ for cold phase, and 
$10^{13}$ cm$^{-3}$ for hot phase, and it is located 
within radius of 10$^{10}$ cm. 

Up to now, there are only a few sources, where the wind density number was 
determined from careful spectral fitting (for example Miller et al. 2013), 
but all of them are showing high values of the wind density around 10$^{16-17}$ cm$^{-3}$.

Such densities are present in upper thermalized disk atmospheres, 
within exactly the same distance from black hole as wind location derived from 
observations. This is a strong argument that the wind can originate from 
upper parts of the disk slabs, and the only question arises how the 
wind is blowing?
There are several mechanisms proposed as: thermal wind, radiation pressure driven 
wind or magnetic wind, but none of these processes can be self consistently 
computed using present computer power.  
Additionally, to obtain a self-consistent model of  radiation pressure
winds it is  critical to include a more detailed
treatment of radiative transfer and ionization in
the next generation of hydrodynamic simulations.

The second argument, showing connection of the wind with accretion disk 
atmospheres is our analyze of \KU\ {\it Suzaku} data. All fitted models were
computed for high density atmospheres and the line profiles 
agree with observations.   
In the wind theory, it is widely accepted that the wind can be launched 
at the accretion disk surface. Upper layers of atmosphere can become 
unstable and start to blow material out  owing to the radiation pressure. 
Our analysis does not contradict this fact, but instead shows that absorption 
at upper atmospheric layers cannot be distinguished from the absorption 
in the wind, which may be launched in the same region.
Data with higher spectral resolution are needed to distinguish between those two 
models. Future satellites with calorimeters, such 
as {\it ASTRO-H} or {\it Athena+},
will yield the answer.

\vspace{1pc}
\section*{Acknowledgements} 
This research was supported by Polish National Science 
Center grant No. 2011/03/B/ST9/03281. It received funding
from the European Union Seventh Framework Program (FP7/2007-2013) 
under grant agreement No.312789.

\section*{References}

\re
Boirin, L., Parmar, A.N., et al. 2004, A\&A, 418, 1061

\re 
D\'iaz Trigo M., Sidoli L., et al. 2012, A\&A, 543, A50 

\re
D\'iaz Trigo M., Boirin L., et al. 2013, Acta Polytechnica, 53, 659

\re
Frank J., King A.R., Lasota J.-P., 1987, A\&A, 178,137 

\re
Hameury J.-M., Lasota J.-P., 2005, A\&A, 443, 283

\re
Korista K.T., Bautista M.A., et al. 2008, ApJ, 688, 108

\re
Krolik J.H., Kriss G.A., 1995, ApJ, 447, 512

\re
Kubota A., Dotani T., et al. 2007, PASJ, 59, 185

\re 
Mauche C.W., Liedahl D.A., et al. 2003, ApJ, 588, L101

\re 
Miller J.M., Raymond J., et al. 2008, ApJ, 680, 1359

\re 
Miller J.M., Raymond J., et al. 2013, arXiv:1306.2915v1

\re 
Pojma\'nski G., 1985, Acta Astronomica, 36, 69 

\re 
Ponti G., Fender R.P. et al. 2012, MNRAS, 422, L11
 
\re 
R\'o\.za\'nska A., Czerny B., et al. 1999, MNRAS, 305, 481

\re 
R\'o\.za\'nska A., Kowalska I., et al. 2008, A\&A, 487, 895

\re 
R\'o\.za\'nska A., Madej J., et al. 2011, A\&A, 527, A47 

\re 
R\'o\.za\'nska A., Madej, J., et al. 2014 A\&A, 562, A81

\re
Shakura N.I., Sunyaev R.A., 1973, A\&A, 24, 337 

\re
Smak J., 1983, Acta Astron., 33, 333

\re 
White N.E., Swank J.H., 1982, ApJ, 253, L61

\label{last}

\end{document}